\documentclass[aps,prl,twocolumn,groupedaddress,showpacs,graphics]{revtex4}
\usepackage{graphicx}

\begin{document}

\title{Terahertz detection schemes based on sequential multi-photon absorption}

\author{Fabrizio Castellano, Rita C. Iotti, and Fausto Rossi}
\affiliation{Dipartimento di Fisica, Politecnico di Torino, Corso
Duca degli Abruzzi 24, 10129 Torino, Italy}

\begin{abstract}

We present modeling and simulation of prototypical multi bound
state quantum well infrared photodetectors and show
that such a detection design may overcome the problems arising when the
operation frequency is pushed down into the far infrared spectral region.
In particular, after a simplified analysis on a parabolic-potential design, we propose a fully three-dimensional model based on
a finite difference solution of the Boltzmann transport equation
for realistic potential profiles. The performances of the proposed
simulated devices are encouraging and support the idea that such
design strategy may face the well-known dark-current problem.

\end{abstract}

\maketitle


Common quantum-well infrared photodetectors (QWIPs) are based on single-bound-state quantum wells, and electronic transitions between such bound state and the continuum
are used for photon detection.~\cite{QWIP} Incident photons excite
bound electrons into the continuum contributing to what is called the
photocurrent, which is the detection signal ``measured'' by such devices.
The well width and composition, and thus the depth of the bound state,
are designed to match with the energy of the photons to be
detected.

QWIPs already demonstrated as successful and reliable devices to
cover the mid-infrared region of the electromagnetic
spectrum.~\cite{QWIP} However the scaling down of the operation
frequency into the terahertz range (1-10 THz) is not
straightforward and still work has to be done after the first
demonstration.~\cite{Liu} The main obstacle is the fact that at
these energies (4.1-41 meV) the dark current, mainly due to the
high-energy tail of the electron distribution function, may become
predominant over the photocurrent signal.

The realization of THz-operating detectors, however, is a crucial
milestone of strategic technological interest. High-performance
quantum-cascade (QC) laser sources operating in this frequency
region have, in fact, already been demonstrated.~\cite{Nature} The
availability of reliable detection strategies is therefore the
necessary step to fully access and exploit the THz-communication
potential. Indeed QC designs have also been demonstrated to
efficiently operate as detectors within a bound-to-bound
scheme.~\cite{QCLdet}

In this paper we present an alternative THz-detector design which,
instead of resorting on a direct bound-to-continuum transition as
conventional QWIPs do, is based on a ladder of equally-spaced
bound states, whose step is tuned to the target frequency of
operation. This kind of photodetector is intrinsically less
responsive to light than the single-bound-state system since many
photons have to be absorbed to excite an electron into the
continuum. However, the reduction of the dark current yields an
overall better signal to noise ratio. The latter is the figure of
merit we choose to characterize this device and depends on the
number $N$ of bound states in the design. The latter is defined as follows
\begin{equation}
 R_{N} = \frac{I^{\rm ph}_{N}}{I^{\rm dark}_{N}}
\end{equation}
where $I^{\rm ph}_{N}$ is the photocurrent and $I^{\rm dark}_N$
the dark current. Since we are focusing on how the number of bound
states may affect the system response, the relevant quantity to
consider is indeed $R_{N}/R_{1}$, that is, the signal to noise
ratio of the system normalized to that of the single-bound-state
QWIP design.

To get a first insight into the general idea of the proposed
architecture, we shall start investigating a simplified strictly one-dimensional device model, i.e., a parabolic quantum well. That latter would
formally produce an infinite sequence of evenly-spaced bound
states along the growth direction. The key approximation is to
limit our analysis to the first $N+M$ levels, the former being the
bound states, the latter representing the continuum. The set of
discrete $M$ wavefunctions is chosen to properly model the real
continuum, i. e., their number is so that the electron population
in the continuum is $M$-independent.
We stress that within this simplified one-dimensional model any in-plane carrier relaxation/thermalization process is necessarily treated via effective/phenomenological inter-level scattering rates. A more quantitative analysis will require a fully three-dimensional treatment of the transport problem (see below).

The evaluation of $R_{N}$ is achieved via a proper rate-equation
modeling of the electron dynamics in our device which allows to
access the charge population $n_j$ of the generic $j$-th level. At
this level of description, we may simply assume that the currents
are proportional to the charge density in the continuum and thus
redefine $R_{N}$ as
\begin{equation}
\label{R}
R_{N} = \frac{\sum_{i=1}^{M}n^{\rm ph}_{i+N}}
{\sum_{i=1}^{M}n^{\rm dark}_{i+N}}
\end{equation}
where the superscripts again refer to the contribution due to the
radiation to be detected (ph) or to the dark current component
(dark). The electron-photon dipole matrix element for in-plane
propagating radiation is exactly known for a parabolic potential
profile and textbook formulas have been used to model the effect
of light in the first $N$ levels.~\cite{Messiah} In the model, the
$M$ continuum levels do not interact with light as it would be in
a real continuum.

The charge populations $n_j$ for a fixed configuration have been derived solving a set of $N+M$ steady-state rate
equations of the form
\begin{equation}
\sum_{i=1}^{N+M}(W_{ij}n_{j} - W_{ji}n_{i}) = 0
\end{equation}
where $W_{ij}$ is the total, i.e. summed over all interaction
mechanisms (electron-phonon, electron-electron, etc.), scattering
rate from state $j$ to state $i$. The contribution to the latter
due to non-optical interactions, $W_{ij}^{\rm dark}$, may be
expressed in terms of a phenomenological lifetime $\tau$,
independent on the level index. In particular, $W_{ij}^{\rm dark}$
is such to achieve thermodynamic equilibrium in the absence
of light according to
\begin{equation}
W_{ij}^{\rm dark} = \frac{1}{(N+M)\tau}
\end{equation}
when $i > j$ (decay) and
\begin{equation}
W_{ij}^{\rm dark} = \frac{e^{-\frac{E_{i} -
E_{j}}{k_{B}T}}}{(N+M)\tau}
\end{equation}
when $i < j$ (excitation).

Figure~1 shows the simulated $R_N/R_1$ ratio, as a function of
$N$, for the parabolic potential profile sketched in the inset and
with a lifetime $\tau = 1$ps. The operation frequency is 3 THz and
the optical power is set in the linear response regime of the
device. The results demonstrate that $R_N/R_1$ increases with the
number of bound states thus suggesting that the proposed detection scheme
actually improves the response with respect to the
single-bound-state QWIP.

Actually, a key-feature of the above parabolic-potential model is
the fact that optical dipole matrix elements increase with the
quantum number, thus intrinsically improving the response as long
as more bound states are considered. This is certainly not the
case for a real potential profile and the feasibility of our
architecture has to be tested on more realistic designs. In
particular, the realistic potential profile of choice in our fully three-dimensional analysis is a triple quantum
well with a wide central well defining the number of bound states
and two thin lateral wells whose dimensions are tailored to tune
the inter-level spacings. This profile can be seen in the inset of
figure~\ref{fig2}.

To get a more refined modeling of the carrier dynamics in our
heterostructure, taking into account the three-dimensional nature of the problem, we adopt a kinetic Boltzmann-like approach.
The fundamental ingredient of the latter is the carrier
distribution function $f_{{\bf k}i}$ which represents the
occupation probability of a single-particle state with in-plane
wavevector {\bf k} in the $i$-th subband. The time evolution of
$f_{{\bf k}i}$ is governed by the following transport equation:
\begin{equation}
\frac{\partial f_{{\bf k}i}}{\partial t} - \frac{e{\bf F}}{\hbar}
\cdot \nabla_{\bf k}f_{{\bf k}i}
 = \sum_{{\bf k}'j} \left(
W_{{\bf k}i,{\bf k}'j}f_{{\bf k}'j} - W_{{\bf k}'j,{\bf
k}i}f_{{\bf k}i} \right)
\end{equation}
where ${\bf F}$ is an external static electric field and $W_{{\bf
k}'j,{\bf k}i}$ is the scattering probability for a process
connecting the state with in-plane wavevector {\bf k} in the
$i$-th subband to the state {\bf k}' in the $j$-th one.

The 3D single-particle electron states forming the basis set of
our scattering dynamics are obtained within the standard
envelope-function approximation, using a plane-wave expansion
strategy as described in Ref.~\onlinecite{BBR}. Dipole matrix elements
are then numerically evaluated while, at this preliminary level of
description, other scattering processes are still modeled in terms
of a phenomenological lifetime. The carrier dynamics then results
from a finite-difference solution of the above Boltzmann transport
equation, in cylindrical coordinates with the polar plane parallel
to the growth direction. In particular, our time-step procedure
starts from a known distribution function, evaluates its
variations over a certain time interval $\Delta t$ and then uses
the new value as the starting point for the forthcoming step.

The simulation is carried out applying a static electric field $F
= 500$ V/cm along the growth direction and then evaluating the
current with and without incident light. The signal-to-noise ratio
$R_N/R_1$ is then estimated according to Eq.~(\ref{R}) and its
values are plotted in figure~\ref{fig2} as a function of the
number of bound states $N$. Our prototypical device is tuned to
operate at 3.3 THz, an electron mean lifetime $\tau = 1$ ps is
assumed and the incident power is kept within the linear response
regime. The graph shows that $R_N/R_1$ grows super-linearly with
the number of bound states $N$, thus confirming the earlier
results obtained with the simplified parabolic-potential model.

This occurs in spite of the fact that the potential profile
sketched in the inset of figure~\ref{fig2} is not the best choice
to optimize the electron-photon interaction. The latter depends on
the overlap integral between the wavefunctions of the initial and
final states of the transition and a proper device design is
crucial, in this respect, for the overall response of the
photodetector. However, at this level of analysis, we focused on
achieving an evenly-spaced set of bound states, since our aim was
to verify whether a multi-photon absorbtion scheme could improve
the detector responsivity over the single bound state design. The
issue of optimizing the device performance in such a novel
architecture will be addressed in a second stage.

To summarize, we have presented a novel QWIP architecture to face
the dark current related problems that affect more conventional
designs especially when the device is operating above cryogenic
temperatures. Our results demonstrate that a multi-bound-state
QWIP structure can significantly improve the signal-to-noise
ratio, thus yielding better results as the number of bound states
is increased. Several aspects, both in the design and the
modeling, deserve a more refined analysis for a quantitative
evaluation and optimization of device performances. These include,
for example, a microscopic description of the relevant scattering
processes (e.g., carrier-phonon), a more optically efficient
potential profile and then simulations of a real device with its
physical boundaries (contacts).







\newpage

\begin{figure}
\includegraphics[width=9cm]{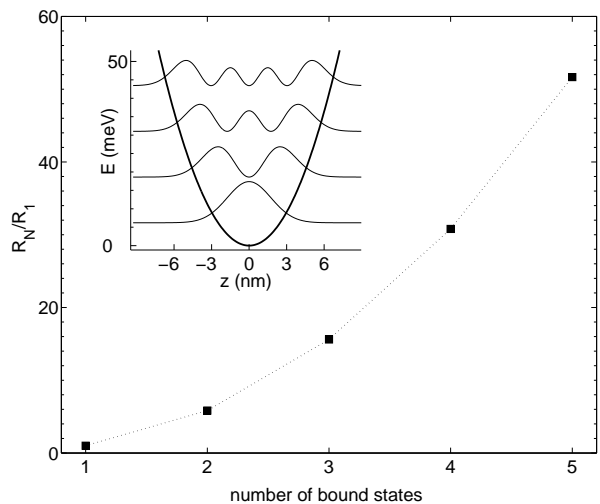}
  \caption{ Normalized signal-to-noise ratio, $R_N/R_1$, for the
  parabolic-potential design (shown in the inset for the case of a
  four-bound-level configuration), as a function of the number of bound states $N$.
  }\label{fig1}
\end{figure}

\begin{figure}
\includegraphics[width=8cm]{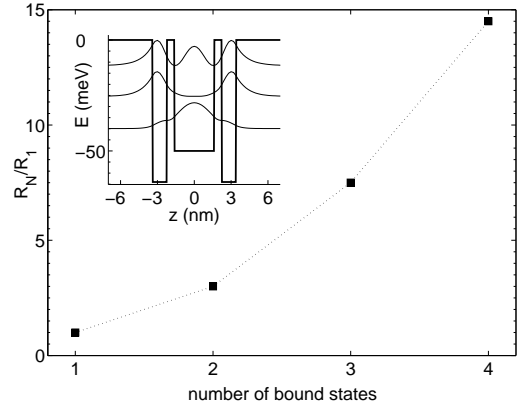}
  \caption{ Normalized signal-to-noise ratio, $R_N/R_1$, for the three-coupled-wells
  potential profile (shown in the inset for the case of a three-bound-level configuration), as
  a function of the number of bound states $N$.
  }\label{fig2}
\end{figure}

\end{document}